\documentclass[a4paper]{article}
\usepackage[latin1]{inputenc}
\sf
\usepackage{amssymb}
\def\R{\mathbb{R}}

\def\C{\mathbb{C}}
\def\H{\mathbb{H}}
\def\Z{\mathbb{Z}}

\usepackage{amssymb}
\def\R{\mathbb{R}}

\def\C{\mathbb{C}}
\def\H{\mathbb{H}}
\def\Z{\mathbb{Z}}

\newtheorem{defi}{Definition}

\begin{document}
\title{A Crossed Module giving the Godbillon-Vey Cocycle } 

\author{Friedrich Wagemann \\
        Institut Girard Desargues -- ESA 5028 du CNRS\\
	Universit\'e Claude Bernard Lyon-I\\
        43, bd du 11 Novembre 1918 \\
	69622 Villeurbanne Cedex FRANCE\\
        tel.: +33.4.72.43.11.90\\
        fax:  +33.4.72.43.00.35\\
        e-mail: wagemann@desargues.univ-lyon1.fr}

\maketitle


\begin{abstract}
We exhibit crossed modules (i.e. certain 4 term exact sequences)
corresponding to the Godbillon-Vey generator for $W_1$, $Vect(S^1)$
and $Vect_{1,0}(\Sigma)$.
\end{abstract} 
 
\section*{Introduction}

There is a well known correspondence in homological algebra between
(equivalence classes of)
exact sequences $\Lambda$-modules, starting in $M$ and ending in $N$
with $n$ modules in between, and elements of
$Ext^{n}_{\Lambda}(N,M)$. For $\Lambda=U(\mathfrak{g})$ the universal
enveloping algebra of a Lie algebra $\mathfrak{g}$, this gives for
example a correspondence between $H^2(\mathfrak{g},V)$ and short eaxct
sequences 

\begin{displaymath}
0\to V\to \hat{g}=V\times\mathfrak{g}\to \mathfrak{g}\to 0
\end{displaymath}

where $\hat{g}$ is the semi-direct product of the abelian Lie algebra
$V$ and $\mathfrak{g}$. The Lie algebra law on  $\hat{g}$ is given by
a 2-cocycle of $\mathfrak{g}$ with values in $V$. Note that the short
exact sequence in uniquely determined on the vector space level by $V$
and $\mathfrak{g}$.

In the same way, there is a correspondence between
$H^3(\mathfrak{g},V)$ and certain 4 term exact sequences called
crossed modules. In these sequences, only the first and the last term
are specified, leaving (du to exactness) a choice of one
$\mathfrak{g}$-module to complete the sequence.

In this article, we exhibit crossed modules corresponding to the
Godbillon-Vey 3-cocycle for $W_1$, $Vect(S^1)$ and the 2-dimensional
analogue of $Vect(S^1)$ $Vect_{0,1}(\Sigma)$ (or
$Vect_{1,0}(\Sigma)$), $\Sigma$ being a compact Riemann surface.

{\bf Acknoledgements}

The author thanks J.-L. Loday for his question stimulating the present
article and C. Kassel for making him aware of reference \cite{LodKas}.
   
\section{Crossed modules}

In the same way as extensions, i.e. 3 term exact sequences, are
related to Lie algebra cohomology in degree 2, crossed modules
\cite{LodKas} \cite{Lod},
i.e. certain 4 term exact sequences, correspond to cohomology in degree 3.

\begin{defi}
A crossed module of a Lie algebra is a homomorphism of
Lie algebras $\mu:\mathfrak{m}\to\mathfrak{n}$ together with a
$\mathfrak{n}$-module structure $\eta$ on $\mathfrak{m}$ such that

(a) $\mu(\eta(n)\cdot m)\,=\,[n,\mu(m)]$ for all $n\in\mathfrak{n}$
and all $m\in\mathfrak{m}$,

(b) $\eta(\mu(m))\cdot m'\,=\,[m,m']$ for all $m,m'\in\mathfrak{m}$.
\end{defi}

One shows that $ker(\mu)\,=:\,V$ is an abelian Lie algebra and that the
action of $\mathfrak{n}$ on $\mathfrak{m}$ induces a structure of a
$\mathfrak{g}\,:=\,coker(\mu)$-module on $V$. Note that in general
$\mathfrak{m}$ and $\mathfrak{n}$ are not $\mathfrak{g}$-modules.

An equivalence of crossed modules is defined in a natural way such
that the restrictions of the maps on the kernel and the cokernel of
$\mu$ are identical. Let us denote by ${\rm crmod}(\mathfrak{g},V)$
the set of equivalence classes of crossed modules. We have then the
fundamental correspondence \cite{LodKas} theorem A.2, p. 138,

\begin{equation} \label{*}
{\rm crmod}(\mathfrak{g},V)\,\cong\,H^3(\mathfrak{g},V).
\end{equation}

Let us briefly review how to associate to a crossed module a 3
cocycle of $\mathfrak{g}$ with values in $V$:

The exact sequence associated to a crossed module reads as follows:

\begin{displaymath}
0\to
V\stackrel{i}{\to}\mathfrak{m}\stackrel{\mu}{\to}\mathfrak{n}\stackrel{\pi}{\to}\mathfrak{g}\to
0
\end{displaymath}

The first step is to take a (continuous) section $\sigma$ of $\pi$ and
to calculate the default of $\sigma$ to be a Lie algebra homomorphism,
i.e.

\begin{displaymath}
\alpha(x_1,x_2):=\sigma([x_1,x_2])-[\sigma(x_1),\sigma(x_2)].
\end{displaymath}

We have obviously $\pi\circ\alpha(x_1,x_2)=0$, because $\pi$ is a Lie
algebra homomorphism, so $\alpha(x_1,x_2)\in im(\mu)=ker(\pi)$. This
means that there is a $\beta(x_1,x_2)\in\mathfrak{m}$ such that

\begin{displaymath}
\mu(\beta(x_1,x_2))\,=\,\alpha(x_1,x_2).
\end{displaymath}

Now, one can easily calculate that $\mu(d\beta(x_1,x_2,x_3))=0$ where
$d$ is the Lie algebra cohomology boundary of cohomology of
$\mathfrak{g}$ with values in $\mathfrak{m}$ where $\mathfrak{g}$ acts
on $\mathfrak{m}$ by $\eta\circ\sigma$.

This means that $d\beta(x_1,x_2,x_3)\in ker(\mu)=im(i)=V$, i.e. there
is a $\gamma(x_1,x_2,x_3)\in V$ such that
$d\beta(x_1,x_2,x_3)=i(\gamma(x_1,x_2,x_3))$. It is fairly obvious
that $\gamma$ is a 3-cocycle of $\mathfrak{g}$ with values in $V$.

\section{A crossed module giving $\theta\in H^3(W_1)$}

In the following, we want to exhibit the 4 term exact sequence related
to the Godbillon-Vey class in $H^3(W_1,\R)$.

Recall that there is a semi-direct product of $W_1$ by its module of
$\lambda$-densities. It is represented by a short exact sequence

\begin{displaymath}
0\to F_{\lambda}\to F_{\lambda}\times W_1\to W_1\to 0,
\end{displaymath}

where the Lie algebra structure on the product is given by

\begin{displaymath}
[(f,a),(g,b)]\,=\,([f,g],L_fb-L_ga+c(f,g)).
\end{displaymath}

Here, $c$ is a 2-cocycle of $W_1$ with values in $F_{\lambda}$, the
module of $\lambda$-densities on the line $\R$. We
take $\lambda = 1$ and 

\begin{displaymath}
c(f,g)\,=\,\left|\begin{array}{cc} f' & g' \\ f'' & g''
\end{array}\right|.
\end{displaymath}

Now consider the short exact sequence of $W_1$-modules

\begin{displaymath}
0\to\R\to F_0\stackrel{d}{\to} F_1\to 0
\end{displaymath}

Since $W_1$ acts by the Lie derivative $L_X = d\circ i_X + i_X\circ d$
on the density modules, the action commutes with the exterior
differential $d$.

We can glue these 2 sequences together to get

\begin{equation}\label{**}
0\to\R\to F_0\stackrel{d}{\to}F_1\times W_1\to W_1\to 0,
\end{equation}

$d$ is trivially a Lie algebra homomorphism from the abelian Lie
algebra $F_0$ to the semidirect product $F_1\times W_1$. There is an
action of $F_1\times W_1$ on $F_0$: it is given by the action of $W_1$
on $F_0$. It gives a structure of a crossed module to the sequence
(\ref{**}). The second compatibility condition is trivial. The first one
reads

\begin{displaymath}
d((f,a)\cdot b)\,=\,[(f,a),db]
\end{displaymath}

The left hand side is just $d(L_fb)=d\circ
i_{f\frac{d}{dt}}\circ d b$. The right hand side gives $(0,L_f(db))$
and commutation of $d$ and $L_f$ shows condition $(a)$. In conclusion,
the 4 term exact sequence gives a crossed module. We claim that the
crossed module corresponds to the Godbillon-Vey class via the
isomorphism (\ref{*}). 

In fact, this is easy: one just has to move up the arrows in the wrong
direction. First, one has to choose a section of the arrow $F_1\times
W_1\to W_1$. But calculating the default of this section to be a 
Lie algebra map just gives the cocycle of the semi-direct
product. Then we have to move up the de Rham differential $d$, i.e. we
have to choose a primitive. Afterwards, we have to take the Lie
algebra coboundary. So, if we denote by a hat the primitive, we have
to calculate 

\begin{displaymath}
d(\hat{c})(f,g,h)\,=\,d(\widehat{\left|\begin{array}{cc} -' & -' \\ -'' & -''
\end{array}\right|})(f,g,h).
\end{displaymath}

But it is obvious that the Lie algebra coboundary commutes with taking
the primitive. Furthermore, it is well known that we have the
following relation

\begin{displaymath}
d(\left|\begin{array}{cc} -' & -' \\ -'' & -''
\end{array}\right|)(f,g,h)\,=\,\left(\left|\begin{array}{ccc} f & g & h \\
f' & g' & h' \\ f'' & g'' & h'' \end{array}\right|\right)'
\end{displaymath}

This expresses that fact that the Gelfand-Fuks cocycle is obtained
from the Godbillon-Vey cocycle by ``integration over the manifold''
(and is easily checked directly).

Thus, we can choose the primitive such that the 3-cocycle associated to
the above 4 term exact sequence is the Godbillon-Vey cocycle which is
the generator of $H^3(W_1,\R)$.\\

{\bf Remarks:}\\

1) This reasoning also shows that $H^2(W_1,F_1)$ which is known to
be 1-dimensional (consequence of Goncharova's theorem, cf \cite{Fuk}
p. 120), is generated by the cocycle

\begin{displaymath}
c(f,g)\,=\,\left|\begin{array}{cc} f' & g' \\ f'' & g''
\end{array}\right|dx.
\end{displaymath}

2) We owe to C. Roger the remark that the crossed module is obtained
under Yoneda product $Ext^1_{U(W_1)}(F_1,\R)\times
Ext^1_{U(W_1)}(\R,F_1)\to Ext^2_{U(W_1)}(\R,\R)\cong
H^3(W_1,\R)$. This possibility bears perhaps deep homological properties. 

\section{A crossed module giving $\theta\in H^3(Vect(S^1),\C)$}

Now we want to generalize the preceeding situation to the
circle. There is an obvious problem to do so: the sequence

\begin{displaymath}
0\to\C\to{\cal F}_0\stackrel{d}{\to}{\cal F}_1\to 0
\end{displaymath}

where ${\cal F}_0={\cal C}^{\infty}(S^1)$ and ${\cal
F}_1=\Omega^1(S^1)$ is not exact, but has a cokernel
$H^1(S^1,\C)=\C$.

In order to get around this problem, let us recall some evident, but
somewhat strange facts:

There is a chain of inclusions of Lie algebras

\begin{displaymath}
W_1^{pol}\hookrightarrow Vect(S^1)\hookrightarrow Vect(\R)
\end{displaymath}

Here, $W_1^{pol}=\bigoplus_{n\geq-1}\C x^{n+1}\frac{d}{dx}$ is the Lie
algebra of (complexified) polynomial vector fields on the line,
$Vect(S^1)=\widehat{\bigoplus_{n\in\Z}\C e^{int}\frac{d}{dt}}$ is the
Lie algebra of complexified vector fields on the circle (the hat
meaning that $e^{int}\frac{d}{dt}$ for $n\in\Z$ is a topological
basis, because periodic functions can be approximated by Fourier
polynomials), and $Vect(\R)=\{f(x)\frac{d}{dx}\,|\,f\in{\cal
C}^{\infty}(\R,\C)\}$ is the Lie algebra of complexified vector fields
on the real line $\R$.

The maps $x^{n+1}\frac{d}{dx}\mapsto -ie^{int}\frac{d}{dt}$ (i.e. one
sets $x=e^{it}$) and $f(t)\frac{d}{dt}\mapsto
\tilde{f}(x)\frac{d}{dx}$ (where the field $f(t)\frac{d}{dt}$ is
lifted to the universal covering to the unique 1-periodic field
$\tilde{f}(x)\frac{d}{dx}$) are easily seen to be Lie algebra homomorphisms.

The strange fact is that $W_1^{pol}$ and
$Vect(\R)$ have the same continuous cohomology, but $Vect(S^1)$ does
not. This is elucidated by the fact that the isomorphism in cohomology
between $W_1^{pol}$ and $Vect(\R)$ is not induced by the above
inclusion, but by the ``Taylor expansion at 0''-map $Vect(\R)\to W_1^{pol}$
(which is a continuous surjection of Fr\'echet spaces by Borel's
lemma). One sees that $W_1^{pol}$ and $Vect(\R)$ correspond to the 2
ways of associating $\R$ to $S^1$: as its tangent space, or as its
universal covering.

I recall all this just to motivate the fact that $Vect(S^1)$ acts on
$F_{\lambda}$ (the $\lambda$-densities on $\R$) and $W_1^{pol}$
acts on ${\cal F}_{\lambda}$ (the $\lambda$-densities on
$S^1$). In the first case, we embed $Vect(S^1)$ into $Vect(\R)$, and
in the second case, we embed $W_1^{pol}$ into $Vect(S^1)$.

This gives us the possibility to consider the exact sequence

\begin{displaymath}
0\to\R\to F_0\stackrel{d}{\to}F_1\times Vect(S^1)\to Vect(S^1)\to 0.
\end{displaymath}

By the same arguments as above, this is a crossed module giving the
Godbillon-Vey cocycle.

\section{A crossed module giving $\theta\in H^3(Vect_{1,0}(\Sigma),\C)$}
 
Now, let us recall that $H^3(Vect_{1,0}(\Sigma),\C)$, where
$Vect_{1,0}(\Sigma)$ is the Lie algebra of differentiable vector
fields of type $(1,0)$ (i.e. locally, such a field is written
$f(z,\bar{z})\frac{\partial}{\partial z}$) on a compact Riemann
surface $\Sigma$, is also of dimension 1 and
generated by a Godbillon-Vey type cocycle, see \cite{Fei},
\cite{Wag1}. This cocycle is the integral over $\Sigma$ over a
determinant as in the usual Godbillon-Vey cocycle, all derivatives
being with respect to $z$.

We propose a crossed module corresponding to this generator.

There are two obvious problems: first of all, there is the same
problem as in the preceeding section, so we have to lift our fields to the
universal covering, and second, there is the problem that the
derivatives involved are only $\partial$, instead of being
$d=\partial+\bar{\partial}$. For the second problem, we define a new
action of $Vect_{1,0}(\Sigma)$ on
$\bigoplus_{i=0,1}\Omega^{i,0}(\Sigma)$ by the modified Cartan formula:

\begin{displaymath}
L_X\omega\,=\,i_X\circ\partial\,+\,\partial\circ i_X.
\end{displaymath}

This means explicitly

\begin{displaymath}
L_X\omega\,=\,\left\{\begin{array}{cc}i_X(\partial(\omega)) & {\rm
if}\,\,\,deg(\omega)=0 \\ \partial(i_X(\omega)) & {\rm
if}\,\,\,deg(\omega)=1\end{array}\right.
\end{displaymath}

One calculates easily that this defines a Lie algebra action  of
$Vect_{1,0}(\Sigma)$ on $\bigoplus_{i=0,1}\Omega^{i,0}(\Sigma)$ and
that 

\begin{displaymath}
0\to\C\to\Omega^{0,0}(\Sigma)\stackrel{\partial}{\to}\Omega^{1,0}(\Sigma)
\end{displaymath}

is an exact sequence of $Vect_{1,0}(\Sigma)$-modules, unfortunately
the last map is not surjective. Now, let us lift these modules to the
universal covering: 

Suppose $g>0$ in order to have as the universal covering of $\Sigma$
either $\C$ or $\H$, and let $U$ denote $\C$ in case the genus is 1 and
$\H$ in case the genus is greater than 1. We have a short exact
sequence of $Vect_{1,0}(\Sigma)$-modules

\begin{displaymath}
0\to\C\to\Omega^{0,0}(U)\stackrel{\partial}{\to}\Omega^{1,0}(U)\to 0
\end{displaymath}

Now we can form the 4 term exact sequence

\begin{displaymath}
0\to\C\to\Omega^{0,0}(U)\stackrel{\partial}{\to}\Omega^{1,0}(U)\times
Vect_{1,0}(\Sigma)\to Vect_{1,0}(\Sigma)\to 0
\end{displaymath}

This sequence gives obviously by the same arguments as before a
crossed module with corresponding 3-cohomology class represented by
the Godbillon-Vey cocycle.

Note that the same construction works for the Lie algebra of
holomorphic vector fields on an open Riemann surface $Hol(\Sigma_r)$
where $\Sigma_r=\Sigma\setminus\{p_1,\ldots,p_r\}$ acting on the
module of holomorphic $\lambda$-densities ${\cal
F}_{\lambda}(\Sigma_r)$ by the modified Cartan formula. This gives a
crossed module describing $H^3(Hol(\Sigma_r),\C)$ which is also known
to be 1-dimensional \cite{Kaw}.

\end{document}